\documentstyle[aps,prb,multicol,epsf]{revtex}
\begin{document}
\draft
\title{Checkerboard local density of states in striped domains pinned by vortices}
\author{Brian M\o ller Andersen, Per Hedeg\aa rd and Henrik Bruus$^*$}
\address{\O rsted Laboratory, Niels Bohr Institute for APG,
Universitetsparken 5, DK-2100 Copenhagen \O\, Denmark\\
$^*$Mikroelektronik Centret, Technical University of Denmark, DK-2800 Lyngby}
\date{December 2002}
\maketitle
\begin{abstract}
We discuss recent elastic neutron scattering and scanning tunneling
experiments on High-T$_c$ cuprates exposed to an applied magnetic field.
Antiferromagnetic vortex cores operating as pinning centers for
surrounding stripes is qualitatively consistent with the neutron data 
provided the stripes have the antiphase modulation. 
Within a Green's function formalism we study the low energy electronic 
structure around the vortices and find that besides the dispersive 
quantum interference there exists a non-dispersive 
checkerboard interference pattern consistent with recent scanning 
tunneling measurements. Thus both experiments can be explained from
the physics of a single CuO$_2$ plane.
\end{abstract}
\begin{multicols}{2}
%\section{Introduction}%
The competing orders in the High-T$_c$ cuprates remain
a strong candidate for explaining some of the unusual features of
these doped Mott insulators\cite{zaanen,emery,kivelson,white,sachdev}. 
The competition between superconducting order and antiferromagnetic
order has recently attracted a large amount of both experimental and
theoretical attention. In particular, experiments in the mixed state
have revealed an interesting coexistence of these
order parameters.\\
Elastic neutron scattering results on La$_{2-x}$Sr$_{x}$CuO$_{2}$ (x=0.10)
have shown that the intensity of the incommensurate peaks in the
superconducting phase is considerably increased when a large magnetic field
is applied perpendicular to the CuO$_{2}$
planes\cite{bella}. This enhanced intensity corresponds to a
spin density periodicity of eight lattice constants $8a_0$ extending far
outside the vortex cores. Similar results
have been obtained for the related material La$_2$CuO$_{4+y}$\cite{kraykovich}.
Nuclear magnetic resonance (NMR) experiments have shown evidence for
antiferromagnetism in and around the vortex cores of near-optimally doped
Tl$_2$Ba$_2$CuO$_{6+\delta}$\cite{kakuyanagi}. Furthermore, muon spin rotation
measurements from the mixed state of YBa$_2$Cu$_3$O$_{6.50}$ find
static antiferromagnetism in the cores\cite{miller}.
Consistent with these findings scanning tunneling microscopy (STM) measurements
performed on the surface of YBa$_2$Cu$_3$O$_{7-\delta}$ and 
Bi$_2$Sr$_2$CaCu$_2$O$_{8+x}$\cite{renner,pan}have revealed very low
DOS inside the vortex cores\cite{andersen,zhu,ghosal}.
Thus, there is increasing evidence for antiferromagnetic correlations
in the vortex cores of the under- and optimally-doped regime of the hole doped cuprates. 
More recent STM measurements of slightly overdoped Bi$_2$Sr$_2$CaCu$_2$O$_{8+x}$ have 
shown a checkerboard halo of the local density of states (LDOS) around the vortex
cores\cite{hoffman1}. This LDOS modulation observed at low energy
$|\omega| < 12$ meV was found to have 
half the period of the spin density wave (SDW) observed by neutron
scattering (i.e. four lattice constants $4a_0$), and to be oriented along the 
crystal axes of the Cu-O plane.\\
The neutron experiments have been analysed within
phenomenological models that assume close proximity to a quantum phase transition
between ordinary superconductivity and a phase with antiferromagnetism
or a phase where superconductivity
coexists with SDW and charge density wave (CDW)
order\cite{sachdev,arovas,demler,polkovnikov}. In these models the
suppression of the superconducting order inside the vortex cores
allows the competing magnetic order to arise. Demler {\sl et al.}\cite{demler} found
that around the vortices the circulating supercurrents can similarly
weaken the superconductivity and induce a
SDW.\\ 
The field-induced checkerboard LDOS pattern in the mixed state has
been recently considered within the framework of several
models\cite{polkovnikov,chenhu,polkovnikov2,chen,franz,zhu_checker}. 
In this paper we add to the discussion by calculating the LDOS in
regions where a d-wave superconductor has been perturbed by induced magnetism.
First, however, we note that a
  checkerboard {\sl spin} modulation is inconsistent with the elastic neutron scattering
experiments by Lake {\sl et al.}\cite{bella} on  La$_{2-x}$Sr$_{x}$CuO$_{2}$
(x=0.10). 
For example, assuming that the checkerboard CDW is intrinsic to the Cu-O planes where it
gives rise to a static SDW checkerboard pattern (Fig. 1a), the
  expected neutron diffraction pattern is shown in Fig. 1b.
\begin{figure}
\centerline{\epsfxsize=\linewidth\epsfbox{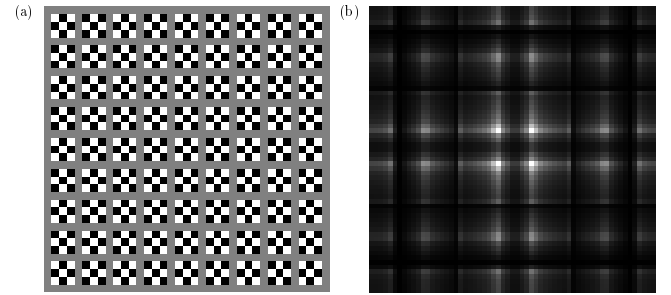}}
\caption{a) Real space picture of the spin structure in a checkerboard
  spin geometry. Black (white) represent spin up (down) while gray reveals
  the superconducting background. In order to simulate the induced
  incommensurability each island of antiferromagnetic spins is out
  of phase with its nearest neighbor. b) Fourier spectrum of the spin 
  checkerboard structure shown in a).} 
\end{figure}
\noindent As is evident there is a 45 degree rotation of the four main
incommensurate peaks and a plaid pattern of the
higher harmonics. The rotated incommensurability (with the correct
absence of an increased signal at ($\pi,\pi$)) shows that this spin structure does not apply to
LSCO for doping levels close to $x=0.10$. It is interesting to note that a
rotation of the incommensurable peaks at low dopings ($x < 0.055$, close
the insulator-superconductor phase transition) has been observed in LSCO\cite{wakimoto}. 
However, there is no simple way to create an antiphase spin
geometry without frustrating the spins
at low dopings where droplets of charge in an antiferromagnetic
background is the expected situation\cite{veilette}.
However, this might be possible in the highly overdoped regime
where the droplets have been inverted to separate magnetic
islands. In that case a 45 degree rotation of the incommensurable
peaks would be consistent with a checkerboard spin pattern. In this light
it would be very interesting to perform an experiment similar to Lake
{\sl et al.}\cite{bella} on highly overdoped LSCO. In the case
of a connected antiferromagnetic background one would also expect a 
large weight at ($\pi,\pi$).\\
The physical picture we have in mind is presented in figure 2a. In
this real space picture an antiferromagnetic core
(center) has pinned a number of surrounding stripes. 
\begin{figure}
\centerline{\epsfxsize=\linewidth\epsfbox{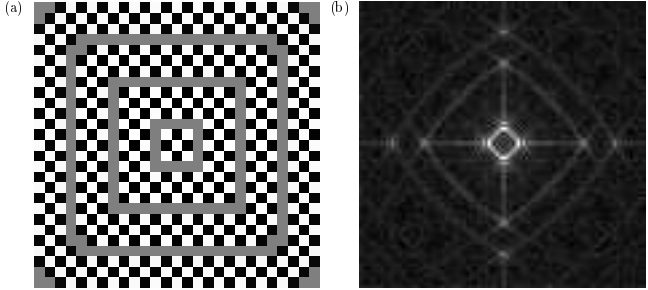}}
\caption{a) The idealized version of a real space spin configuration
  consistent with our physical picture. b) Fourier spectrum of the spin
  density order from a). Almost all the induced weight is located in
  the four incommensurable peaks.}  
\end{figure}
\noindent This pinning effect of SDW by magnetic vortex cores is a well-known
effect from numerical studies\cite{chen}.\\
Both experimentally\cite{tranquada} and
theoretically\cite{zaanen,kivelson,sachdev,zachar} 
we expect an antiphase modulation of the induced antiferromagnetic ring domains.
Indeed as seen in Figure 2b the related diffraction pattern is
qualitatively consistent with measurements by Lake {\sl
  et al.}\cite{bella} of enhanced intensity at the incommensurate
points.\\  
Without an applied magnetic field, only disorder can produce a
similar pinning effect of the fluctuating stripes\cite{kapitulnik}. 
In addition to the creation of more pinning centers when applying a magnetic field, the
single site impurities are expected to pin much weaker than the large
``impurities'' created by the flux lines. This is qualitatively
consistent with the measurements by Lake {\sl et al.}\cite{bella} of
the temperature dependence of the increased magnetic signal for
different magnetic field strengths.\\
This leads to the question of the electronic structure around
extended magnetic perturbations in d-wave superconductors. 
The many experiments indicating
coexistence of d-wave superconductivity and antiferromagnetism
mentioned above motivate studies of simple models that enable one to
calculate the LDOS in such regions.\\
The model Hamiltonian defined on a 2D lattice is given by
\begin{eqnarray}\label{hamil0}
\hat{H}^0 &=&- \sum_{\left< n,m \right>\sigma}t_{nm}\hat{c}^\dagger_{n\sigma} \hat{c}_{m\sigma} -\mu \sum_{n\sigma}
\hat{c}^\dagger_{n\sigma} \hat{c}_{n\sigma} \\ \nonumber
&+& \sum_{\left< n,m \right>} \left( \Delta_{n,m}
\hat{c}^\dagger_{n\uparrow} \hat{c}^\dagger_{m\downarrow} +H.c.\right)
\end{eqnarray}
\begin{equation}\label{hamilaf}
\hat{H}^{int}= \sum_{n} M_n
\left(\hat{c}^\dagger_{n\uparrow} \hat{c}_{n\uparrow} - \hat{c}^\dagger_{n\downarrow}
  \hat{c}_{n\downarrow} \right)
\end{equation}
where $\hat{c}^\dagger_{n\sigma}$ creates an electron with spin $\sigma$ at site
$n$ and $\mu$ is the chemical potential. The staggering is included in
$M_n=\left( -1 \right)^{\mathbf{n}} M$. The strength of the
antiferromagnetic and superconducting coupling is given by $M$ and
$\Delta$, respectively.\\
The Hamiltonian $\hat{H}^0+\hat{H}^{int}$ is a simple mean-field
lattice model to describe the phenomenology of the coexistence of 
d-wave superconducting and antiferromagnetic
regions. This approach is similar to the starting point of many recent
Bogoliubov-de Gennes calculations\cite{andersen,zhu,chen}. The Hamiltonian in
Eqn. (\ref{hamil0})-(\ref{hamilaf}) can be viewed as the  mean-field
Hamiltonian of a $t-U-V$ Hubbard model, where the nearest neighbor
attraction $V$ gives rise to the d-wave superconductivity. In contrast the
on-site Coulomb repulsion $U$ only causes the antiferromagnetism. In this 
article we do not diagonalize $\hat{H}$ in the Bogoliubov-de Gennes scheme since such
lattice calculations require unrealistically large gap $\Delta$ and
magnetic field values. Instead we solve the
Dyson equation exactly by inverting a large matrix. This approach has
previously been utilized extensively to study various short-ranged impurity effects in
superconductors\cite{flatte}, but can also be used for extended
perturbations embedded in a $\hat{G}_0$ medium. Here $\hat{G}_0$ is the Green's
function of the parent medium, in this case a 
d-wave BCS superconductor. This Green's function is given by
\begin{equation}\label{defG0}
\hat{G}_0^{-1}({\mathbf{p}},\omega)= (\omega+i \delta)\tau_0 -
  \xi_{\mathbf{p}}\tau_3-\Delta_{\mathbf{p}} \tau_1
\end{equation}
where $\tau_\nu$ denote the Pauli matrices in Nambu space and the gap
function $\Delta_{\mathbf{p}}=\frac{\Delta_0}{2} \left( \cos \left( p_x \right) - \cos \left( p_y \right) \right)$.
The lattice constant $a_0$ is set to unity and $\xi_{\mathbf{p}}=\epsilon_{\mathbf{p}}-\mu$ with
\begin{equation}
\epsilon_{\mathbf{p}}=-2 t \left( \cos \left( p_x \right)\!+\cos \left( p_y \right)
\right)\!-\! 4 t' \left( \cos \left( p_x \right) \cos \left( p_y \right)
\right).
\end{equation}
Here $t(t')$ refers to the nearest (next-nearest) neighbor hopping
integral and $\mu$ is the chemical potential.
We perform the 2D Fourier transform of $G_0(\mathbf{p},\omega)$ numerically
by utilizing a real space lattice of $1000 \times 1000$ sites and a
quasiparticle energy broadning of $\delta=1.0 \mbox{meV}$.\\
To simulate the situation around optimal doping of the hole doped cuprates the following
parameters are chosen: $t=300 \mbox{meV}$, $t'=-120 \mbox{meV}$,
$\Delta_0=25 \mbox{meV}$, $\mu=-354 \mbox{meV}$.
When the real space domain affected by $H^{int}$ involves a finite
number of lattice sites $N \times N$ we can solve
the Dyson equation exactly to find the full Greens function.
Writing the Dyson equation in terms of real-space (and Nambu) matrices it
becomes
\begin{equation} 
\underline{\underline{G}}(\omega)=\underline{\underline{G}}^0(\omega)
\left( \underline{\underline{1}} - \underline{\underline{H}}^{int}
  \underline{\underline{G}}^0(\omega) \right)^{-1}.
\end{equation}
The size of the matrix $\left( \underline{\underline{1}} - \underline{\underline{H}}^{int}
  \underline{\underline{G}}^0(\omega) \right)$ is given by $(d \times N^2) \times (d \times N^2)$
where $d$ is an integer equal to the number of components in the Nambu particle-hole spinor and 
$N$ denotes the total number of lattice sites affected by the magnetic
  perturbation. Therefore a
real-space lattice with $25 \times 25$ sites affected by perturbations
  results in a $(1250 \times 1250)$ matrix to be inverted.\\ 
Knowing the full Greens function we obtain the LDOS
  $\rho({\mathbf{r}},\omega)= -\frac{1}{\pi} \mbox{Im} \left[
  G_{11}({\mathbf{r}},\omega) +  G_{22}({\mathbf{r}},-\omega) \right]$,   
which is proportional to the differential conductance measured in the
STM experiments.\\
We have checked that the above approach reproduces the expected LDOS
for unitary non-magnetic impurities in d-wave
superconductors\cite{rosengren}. Also in this one-impurity case we
reproduce the constant-energy LDOS maps recently
calculated by Wang {\sl et al.}\cite{dhlee,comment1}\\
Motivated by the qualitative agreement of the spin structure in
figure 2a with the neutron data, we assume that this
represents the induced magnetism around the vortices and calculate the
LDOS in this striped environment.
To this end we simply restrict the sum in Eqn. (\ref{hamilaf}) to include the sites
within these magnetic regions. The system is depicted in figure 2a where
the grey background reveals the underlying superconducting
state. Again the
black (white) squares correspond to the sites affected by the
staggered magnetic perturbation.\\
Figures 3 and 4 show real-space maps of the LDOS summed over a small energy window from -12 meV to +12 meV in
intervals of 1meV for different strengths of the antiferromagnetic
perturbation $M$. The vortex center is located in the center of the
images. Figure 3 (4) is calculated with (without) the antiphase
modulation of the adjacent stripes. Thus the spin configuration of figure 2a corresponds to the
images in figure 3. The clear difference between the LDOS images of figures 3 and 4 reveals
that the STM technique can be used to determine this
phase relation. It is clearly seen from both figures 3 and 4 that the
low energy LDOS structure eventually becomes ringshaped as the
magnitude of $M$ increases. In this limit the pinned stripes 
operate as steep potential walls. Figures 3a and 3b seem to display the closest resemblence to the
experimental data\cite{hoffman1} which indicates that the induced
magnetism is very weak. In figure 5 we show the Fourier transform of
several constant energy LDOS images for $M$ = 100 meV with
the antiphase spin modulation included. In these figures the Fourier component
$\mathbf{q}=0$ is located at the center. The detailed energy
dependence of these images is caused by quasiparticle
interference effects as pointed out by Wang {\sl et al.}\cite{dhlee}
in the case of a single impurity. 
\begin{figure}
\centerline{\epsfxsize=\linewidth\epsfbox{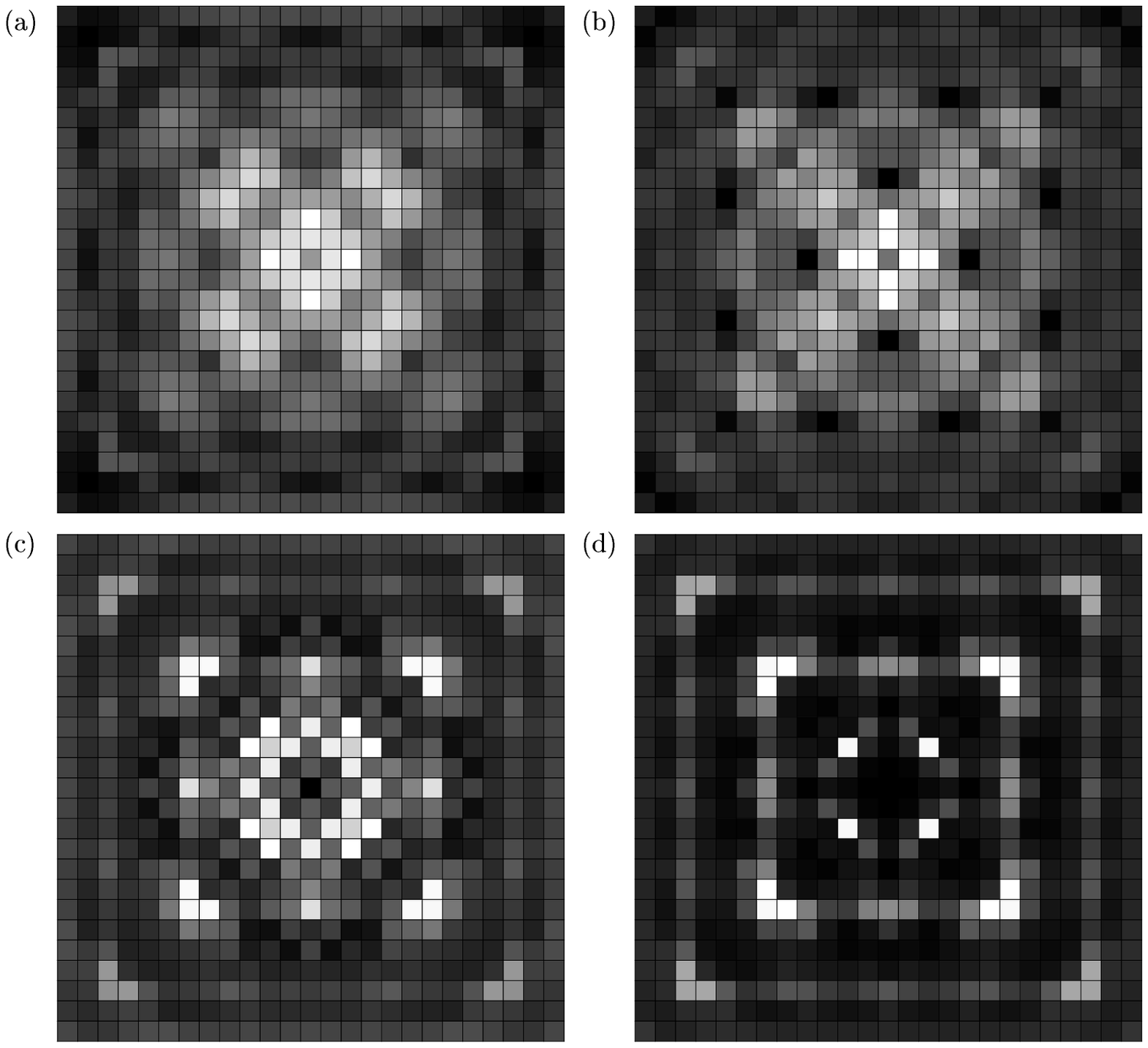}}
\caption{Real-space LDOS summed from -12 meV to +12 meV for: a) M = 35 meV, 
  b) M = 100 meV, c) M = 200 meV, d) M = 300 meV.}  
\end{figure}
\noindent The dispersive features of the images presented in
figure 5 are dependent on the microscopic parameters and the
associated Fermi surface. However, it is also evident that the
ringshaped stripes surrounding the vortex cores give rise to
non-dispersive intensity around ${\mathbf{q}}=\frac{2\pi}{a_0}(\pm 1/4,0)$ and
${\mathbf{q}}=\frac{2\pi}{a_0}(0,\pm 1/4)$. This in turn leads to the checkerboard pattern in
the low energy sums of the LDOS displayed in figures 3 and 4 whereas
the dispersive features fade away in these summed LDOS images\cite{kapitulnik}.
\begin{figure}
\centerline{\epsfxsize=\linewidth\epsfbox{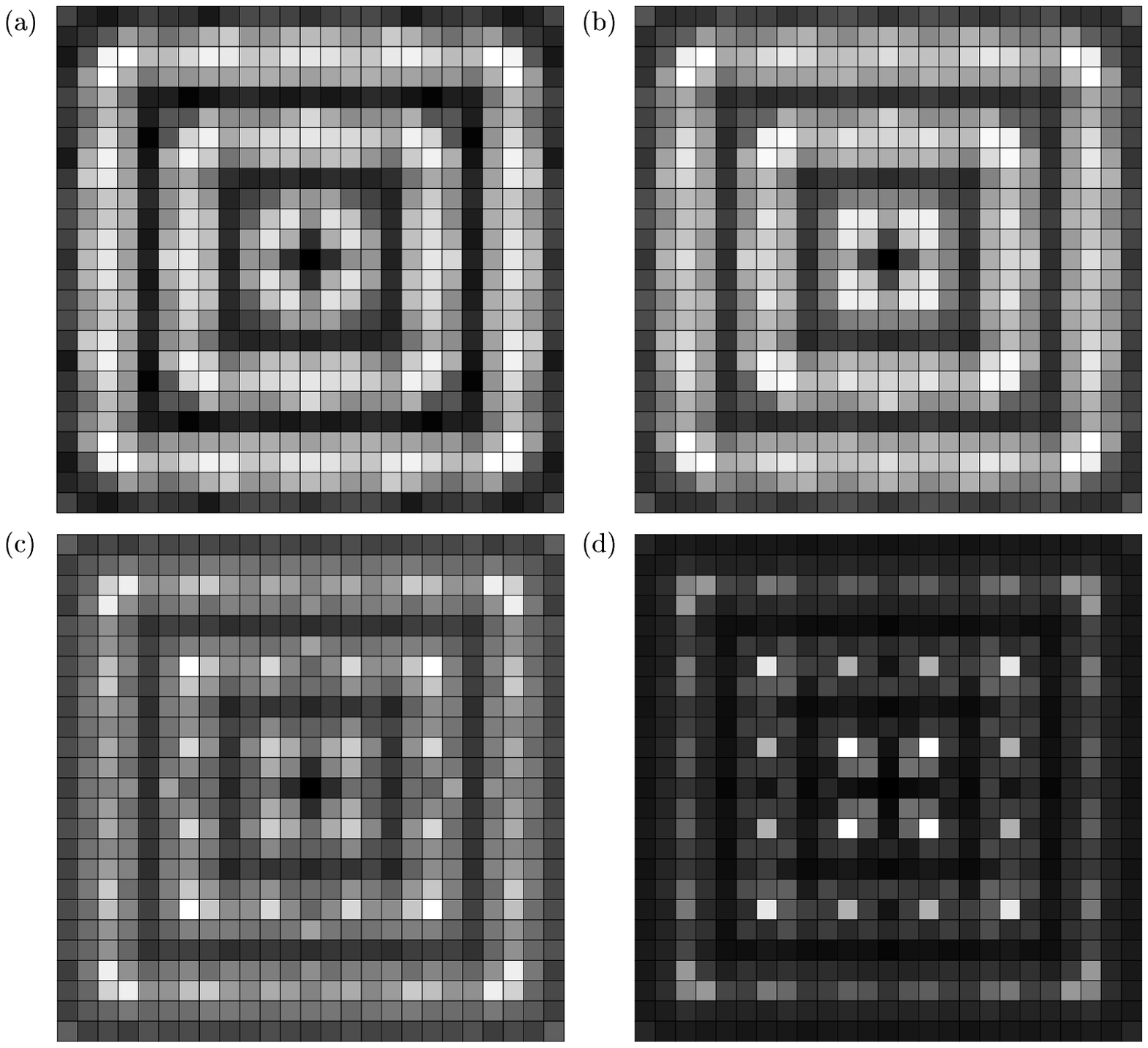}}
\caption{Real-space LDOS summed from -12 meV to +12 meV for: a) M = 35 meV, 
  b) M = 100 meV, c) M = 200 meV, d) M = 300 meV.}  
\end{figure}
\noindent We have confirmed this fact by identifying similar non-dispersive features in
the LDOS around configurations with different periodicities. For
instance a structure with $2a_0$ charge
periodicity leads to non-dispersive intensity around ${\mathbf{q}}=\frac{2\pi}{a_0}(\pm 1/2,0)$ and
${\mathbf{q}}=\frac{2\pi}{a_0}(0,\pm 1/2)$.
\begin{figure}
\centerline{\epsfxsize=\linewidth\epsfbox{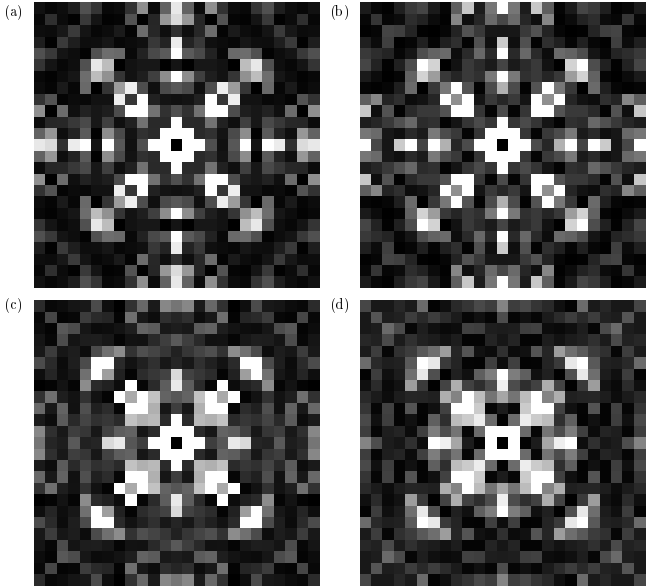}}
\caption{Fourier images of the constant energy LDOS maps for M = 100
  meV and a) $\omega$ = 3 meV, 
  b) $\omega$ = 6 meV, c) $\omega$ = 9 meV, d) $\omega$ = 12 meV.}  
\end{figure}
\noindent In the above calculation we have not yet included the Doppler shift from the circulating
supercurrents or the gap suppression close to the vortex core. As pointed out by
Polkovnikov {\sl et al.}\cite{polkovnikov} the former effect is not
expected to produce significant changes of the four-period
modulations. As for the latter we have
checked that a gap suppression only leads to minor quantitative
changes in the dispersive part of the LDOS. Finally, Podolsky {\sl
  et al.}\cite{podolsky} discussed scenarios of weak translational
symmetry breaking and found that in order to explain quantitatively the
{\sl zero-field} STM results by Howald {\sl et al.}\cite{kapitulnik} one
needs to include dimerization, the modulation of the electron hopping. 
This dimerization will also produce quantitative changes, but not
alter the qualitative conclusion that pinned stripes produce checkerboard LDOS.\\
%\section{Summary}%
In summary we have discussed the phenomenology of a simple physical picture of
pinned stripes around vortex cores which are forced to be
antiferromagnetic by an applied magnetic field. The induction of
magnetic striped race-tracks around the core is consistent with the neutron diffraction spectra
observed on LSCO with a doping level near x=0.10. As expected this is
only true if the stripes are out of phase with their neighbors in the usual sense.
In materials where a checkerboard spin pattern is relevant (possibly
Bi2212 or overdoped LSCO), we show that a 45 degree rotation of the
main incommensurable peaks is to be expected. Finally we studied the
electronic structure around the vortices and identified a
non-dispersive feature in the LDOS arising from the induced static
antiferromagnetism. This feature gives rise to the checkerboard LDOS 
observed experimentally by Hoffman {\sl et al.}\cite{hoffman1} Thus
both the STM measurements and the enhanced intensity of the
incommensurable peaks observed by neutron diffraction can be
ascribed to the phenomena of a single CuO$_2$ plane.\\
Acknowledgement: Support by the Danish Natural Science Research
Council, Ole R\o mer Grant No.\ 9600548.

\end{multicols}
\end{document}